\documentclass[twocolumn,showpacs,preprintnumbers,amsmath,aps,amssymb]{revtex4}
\usepackage{graphicx}% Include figure files
\usepackage{dcolumn}% Align table columns on decimal point
\usepackage{bm}% bold math

\begin{document}

\preprint{}

\title{Scalar potentials out of canonical quantum cosmology}
\author{W. Guzm\'an}
 \email{wguzman@fisica.ugto.mx}
\author{M. Sabido}%
 \email{msabido@fisica.ugto.mx}
\author{J. Socorro}%
 \email{socorro@fisica.ugto.mx}
\author{L. Arturo Ure\~na-L\'opez}
 \email{lurena@fisica.ugto.mx}
\affiliation{Instituto de F\'{\i}sica de la Universidad de Guanajuato,\\
 A.P. E-143, C.P. 37150, Le\'on, Guanajuato, M\'exico}%

\date{\today}% It is always \today, today,

\begin{abstract}
Using canonical quantization of a flat FRW  cosmological model
containing a real scalar field $\phi$ endowed with a scalar potential
$V(\phi)$, we are able to obtain exact and semiclassical solutions of
the so called Wheeler-DeWitt equation for a particular family of
scalar potentials. Some features of the solutions and their
classical limit are discussed.
\end{abstract}

%\pacs{02.30.Jr; 04.20.Jb; 04.60.Ds; 04.60.Kz; 98.80.Jk; 98.80.Qc.}
\maketitle

It is a common issue in Cosmology nowadays to make use of scalar
fields $\phi$ as the responsible agents of some of the most intriguing
aspects of our universe. Just to mention a few, we find scalar
fields in inflaton (as seeds of the primordial perturbations for
structure formation); in cold dark matter models of the formation
of the actual cosmological structure, and also in dark energy models
that intend to explain the current accelerated expansion of the
universe
\cite{Ratra:1987rm,Chimento:1995da,Copeland:1997et,Lyth:2000,SW,tmatos3,tmatos4,Arbey:2001qi,tmatos5,tmatos6,tmatos8,Peebles:2002gy,Alam:2004jy,Sahni:2004ai,Maartens:2002pg,Dimopoulos:2003ss,max1}.

The key feature for such a flexibility of scalar fields
(spin-$0$ bosons) is the freedom one has to propose a \textit{scalar
  potential} $V(\phi)$, which encodes in itself the (non
gravitational) self-interactions among the scalar particles. The
literature on scalar potentials is enourmosly vast, and most of the
recent papers are aimed to explain the SnIa results that suggest the
existence of dark energy \cite{Ratra:1987rm,Chimento:1995da,Copeland:1997et,SW,Urena-Lopez:2000aj,Peebles:2002gy,Alam:2004jy,Sahni:2004ai}.

Recently, scalar fields coupled to gravity (in a FRW background) have
also appeared in connection to the so called string theory
landscape \cite{landscape,landscape1,landscape2,max1}, where the scalar
potential $V(\phi)$ is usually thought as having many valleys, which
represent the different vacua solutions. The hope is that the
statistics of these vacua could explain, for example, the smallness
the cosmological constant (the simplest candidate for dark energy).

Scalar fields also appear in the study of tachyon dynamics. For
instance, in the unstable D-brane scenario, the scalar potential in
the tachyon effective action around the minimum of the potential is of
the form $V(\phi)=e^{-\alpha
  \phi/2}$\cite{Sen:2002an,Sen:2002qa}. Currently, there has been a
lot of interest in the study of tachyon driven cosmology
\cite{Gorini:2003wa,Garcia-Compean:2005zn}.

On the other hand, scalar fields have also been used within the so
called (canonical) Quantum Cosmology (QC) formalism, which deals with a
very early quantum epoch of the cosmos. Again, scalar fields act as
matter sources, and then play an important role in determining the
evolution of such an early universe.

QC means the quantization of minisuperspace models, in which the
gravitational and matter variables have been reduced to a finite number of
degrees of freedom. These models were extensively studied by means of
Hamiltonian methods in the 1970s (for reviews see
\cite{ryan,maccallum}). It was first remarked by Kodama
\cite{kodama1,kodama2}, that solutions to the Wheeler-DeWitt equation
(WDW) in the formulation of Arnowitt, Deser and Misner (ADM) are related to
Ashtekar formalism (in the connection representation) by
$\Psi_{ADM}=\Psi_A e^{\pm i\Phi_A}$, where $\Phi_A$ is the homogeneous
specialization of the generating functional \cite{ashtekar} of the
canonical transformation from the ADM variables to Ashtekar's. This
function was calculated explicitly for the diagonal Bianchi type IX
model by Kodama, who also found $\Psi_A=constant$ as solution. Since
$\Phi_A$ is purely imaginary, for a certain factor ordering, one
expects a solution of the form $\Psi= W e^{\pm \Phi}$.

Our aim in this paper is to determine which scalar potentials can
arise as exact solutions to the WDW equation of QC, as well as which
of them can be valid at the semiclassical level. For this we will use
some of the ideas presented in the previous paragraphs to derive the WDW
equation and find exact solutions to the quantum cosmological
model. Though there are many solutions in principle, we will focus on
those that may be relevant solutions for the early universe.

We begin by writing down the line element for a homogeneous and isotropic
universe, the so called Friedmann-Robertson-Walker (FRW) metric,
in the form
\begin{equation}
ds^2= -N^2(t) dt^2 + a^2(t)\left[\frac{dr^2}{1-k r^2}
+r^2 d \Omega^2 \right] \, ,
\end{equation}
where $a(t)$ is the scale factor, $N(t)$ is the lapse function, and
$k$ is the curvature constant that takes the values $0,+1,-1$, which
correspond to a flat, a closed and an open universe, respectively. The
effective action we are going to work on is \footnote{This action
  appears in connection to the string theory landscape, see for
  example\cite{Gorini:2003wa,landscape}.}
\begin{equation}
S_{tot}=S_g+S_{\phi}=\int  dx^4 \sqrt{-g} \left[ R-2 \Lambda -
 \frac{\dot{\phi}^2}{2}+V(\phi) \right] \, ,
\label{accion}
\end{equation}
where $\phi$ is a scalar field endowed with a scalar potential
$V(\phi)$, and $\Lambda$ is a cosmological constant \footnote{We are
  taking units such that $c=(8\pi G/3)=\hbar=1$.}. The Lagrangian
for a flat FRW cosmological model is
\begin{equation}
{\cal L}=\frac{1}{2} \frac{a \dot{a}^2}{N} - \frac{1}{2} a^3\Lambda N-
\frac{a^3}{2} \frac{\dot{\phi^2}}{N}+V(\phi)a^3N \, , \label{lag}
\end{equation}
and then the canonical momenta are found to be
\begin{subequations}
\label{pps}
\begin{eqnarray}
P_a= \frac {\partial L}{\partial \dot{a} }=\frac{a\dot{a}}{N} \, ,
\qquad \dot{a}=\frac{NP_a}{a} \, , \label{pa}\\
P_\phi= \frac {\partial L}{\partial \dot{\phi} }=-\frac{a^3\dot{\phi}}
{N} \, , \qquad\dot{\phi}=-\frac{NP_\phi}{a^3} \, .\label{pp}
\end{eqnarray}
\end{subequations}
We are now in position to write the corresponding Hamiltonian
\begin{equation}
{\cal H}=\frac{1}{2}
\frac{P_a^2}{a}-\frac{1}{2}\frac{P_\phi^2}{a^3}-a^3 V(\phi,\Lambda)
\, , \label{uno}
\end{equation}
where we have written $V(\phi,\Lambda)=2V(\phi) -\Lambda $.

The WDW equation for this model is achieved by replacing $P_{q^\mu}$
by $-i \partial_{q^\mu}$ in Eq.~(\ref{uno}) with $q^\mu=(a,\phi)$.
We now perform the change of variable $a=e^\alpha$\footnote{Usually,
  $\alpha$ is called the number of $e$-foldings, and then
  $\dot{\alpha}=H$, where $H$ is called the Hubble parameter.}; hence
the total Hamiltonian can be written as
\begin{equation}
{\cal H}=\frac{e^{-3\alpha}}{2} \left [-\frac{\partial^2}{\partial\alpha^2} +
\frac{\partial}{\partial\alpha} +\frac{\partial^2}{\partial\phi^2}  -
e^{6\alpha} V(\phi,\Lambda)\right]=0 \, . \label{ham}
\end{equation}

Following the suggestion by Hartle and Hawking \cite{haha} we do a
  \textit{semi-general factor ordering} on $e^{-3\alpha}$ and
  $P_\alpha$, and then
\begin{equation}
- e^{-(3- q)\alpha}\, \partial_\alpha e^{-q\alpha} \partial_\alpha =- e^{-3\alpha}\, \partial^2_\alpha + q\, e^{-3\alpha} \partial_\alpha  \, ,
\end{equation}
where $q$ is any real constant. Under this factor ordering the WDW reads
\begin{equation}
\Box \, \Psi + Q \frac{\partial \Psi}{\partial  \alpha} - e^{6\alpha}
V(\phi,\Lambda) \Psi =0 \, , 
\label {WDW}
\end{equation}
where $Q=q+1$, $\Psi$ is called the wave function of the universe,
and $\Box \equiv -\partial^2_\alpha + \partial^2_\phi$ is the two
dimensional d'Alambertian operator.

Before going further, we would like to mention here the so called
\textit{semiclassical} limit of the WDW equation. This is achieved by
taking $\Psi = e^{-S}$, and imposing the usual WKB conditions on $S$
\begin{equation}
\left( \frac{\partial S}{\partial \alpha}\right)^2 \gg \left| 
\frac{\partial^2S}{\partial \alpha^2}\right| ,
\qquad \quad
\left( \frac{\partial S}{\partial \phi }\right)^2 \gg \left| 
\frac{\partial^2 S}{\partial \phi^2}\right| \, . 
\label{condiciones}
\end{equation}
Hence, the WDW equation, under a particular factor ordering ($Q=0$), becomes
what is called the Einstein-Hamilton-Jacobi (EHJ) equation,
\begin{equation}
(\nabla S)^2 - U = 0 \, . \label{hj} 
\end{equation}
This equation is also obtained if we introduce the following
transformation on the canonical momenta $P_{q^\mu}\rightarrow
\frac{\partial S}{\partial q^\mu}$ in equation (\ref{ham}). In
consequence, ${\cal H}=0$ is equivalent to the known Friedmann equation, and
Eqs.~(\ref{pps}) provide the classical solutions of the
Einstein-Klein-Gordon equations.

Eq.~(\ref{WDW}) resembles the equation of a damped massive wave
equation, where the mass term is provided by the scalar
potential. Thus, it is interesting to note that, for a free scalar
field ($V(\phi)=0$) and $Q=0$, the wave function of the universe
consists of two wave solutions traveling along the directions $\alpha
\pm \phi$, which are indeed the \textit{classical trajectories} on the
$\{\alpha,\phi\}$ plane.

Taking the following ansatz for the wave function\cite{MR,OS}
\begin{equation}
\Psi(\alpha,\phi) = W(\alpha,\phi) e^{- S(\alpha,\phi)} \, \label{wavefunction}
\end{equation}
where $S(\alpha,\phi)$ is now termed as the \textit{superpotential function},
Eq.~(\ref{WDW}) can be splitted into the EHJ equation~(\ref{hj}) for
$S$, and into the following equations
\begin{subequations}
\label{WDWa}
\begin{eqnarray}
W \left( {\square \, S} + Q \frac{\partial S}{\partial \alpha} \right)
+ 2 {\nabla \, W \cdot \nabla \, S} &=& 0 \, , \label{cons} \\
{\square \, W} + Q\frac{\partial W}{\partial \alpha}&=&0 \, , \label{wdwho}
\end{eqnarray}
\end{subequations}
with ${\nabla \, W}\cdot {\nabla \, S} \equiv -\left( \partial_\alpha
W \right) \left(\partial_\alpha S \right) + \left( \partial_\phi
W \right) \left( \partial_\phi S \right)$, $(\nabla)^2 \equiv
-\left( \partial_\alpha \right)^2 +\left( \partial_\phi \right)^2$,
and $U(\phi,\Lambda)=e^{6\alpha} V(\phi,\Lambda)$. 

In order to find solutions of the WDW equation, we will choose to solve
Eqs.~(\ref{hj}) and~(\ref{cons}), whose solutions will have to comply
with Eq.~(\ref{wdwho}), which will be our \textit{constraint equation}.

Let us start with Eq.~(\ref{hj}), which is an equation for the
superpotential function only. If
$S(\alpha,\phi)=(1/\mu)e^{3\alpha}g(\phi)$, then Eq.~(\ref{hj})
becomes an ordinary differential equation for $g(\phi)$ in terms of
the scalar potential as
\begin{equation}
\left( \frac{d g}{d \phi} \right)^2 - 9 g^2  = \mu^2 V(\phi,\Lambda) \,
. \label{separationa}
\end{equation}

This last equation has several exact solutions, which can be generated
in the following way. Let us consider that $V(\phi,\Lambda)=g^2 F(g)$,
where $F(g)$ is an arbitrary function of its argument. Thus,
Eq.~(\ref{separationa}) can be written in quadratures as
\begin{equation}
\Delta \phi = \int \frac{d \ln g}{\sqrt{9+\mu^2 F(g)}} \, . \label{gquad}
\end{equation}
From Eq.~(\ref{gquad}) we can solve for $g$ as a function of $\phi$,
and then find the corresponding scalar potential that leads to an
exact solution of the EHJ. Some solutions for the
scalar potential are shown in Table~\ref{t:solutions}.
\begin{table}%[H]
\caption{ \label{t:solutions} Some exact solutions of
  Eq.~(\ref{gquad}) and their corresponding scalar potentials. Here,
  $V_0$ is an arbitrary constant, and $n$ is any real number.}
\begin{ruledtabular}
\begin{tabular}{cc}
$F(g)$ & $V(\phi,\Lambda)$ \\ \hline 
$V_0$ & $V_0 \exp \left(- 2 B \Delta \phi \right) \, , \, B =
  \sqrt{9+\mu^2 V_0}$ \\
$g^{-n}$ & $\left\{ (\mu^2/9) \left[ \cosh^2 \left( \frac{3n}{2}\Delta
    \phi \right) -1 \right] \right\}^{(2-n)/n}$ \\
$\ln g$ & $u e^{2u} \, , \, u=(\mu \Delta \phi/2)^2 - (3/\mu)^2$ \\
$(\ln g)^2$ & $u^2 e^{2 u} \, , \, u=(3/\mu) \sinh \left( \mu
  \Delta \phi \right)$
\end{tabular}
\end{ruledtabular}
\end{table}

Next, we assume that $W=e^{\left[ u(\alpha)+v(\phi) \right]}$ in
Eq.~(\ref{cons}); after a bit of algebra, we obtain
\begin{equation}
W = \exp \left\{ \frac{k}{2} \left[ \frac{\alpha}{3}+\int
  \frac{d\phi}{\partial_\phi (\ln g)} \right] +\frac{Q}{2}\alpha- \frac{\mu^2}{4} \int
  \frac{d[V(\phi)]}{\left( \partial_\phi g \right)^2} \right\} \,
  , \label{separation1}
\end{equation}
where $k$ is an arbitrary constant. One only needs to verify under
which conditions solutions in Eqs.~(\ref{separationa})
and~(\ref{separation1}) comply with the constraint
equation~(\ref{wdwho}), which takes the following form
\begin{subequations}
\begin{eqnarray}
\partial^2_\phi v &+& \left(\partial_\phi v \right)^2 -
\frac{k^2-9Q^2}{36} = 0 \, , \\
\partial_\phi v &=& \frac{k}{2} \frac{1}{\partial_\phi (\ln g)}
 - \frac{\mu^2}{4} \frac{\partial_\phi \left[V(\phi)\right] }{\left(
   \partial_\phi g \right)^2} \, . \label{wdwho2}
\end{eqnarray}
\end{subequations}
It is clear that the constraint equation is not easy to satisfy;
but it can be done approximately as we shall show below.

For completeness, the \textit{classical} solutions of the EHJ equation
arising from Eq.~(\ref{separationa}) are given by
\begin{subequations}
\label{classical}
\begin{eqnarray}
\frac{\alpha}{3} + \int \frac{d\phi}{\partial_\phi (\ln g)}
&=& \textrm{const.} \, , \label{classicala} \\
-\mu \int \frac{d\phi}{\partial_\phi g} &=& \Delta t \, . \label{classicalb}
\end{eqnarray}
\end{subequations}

We will now focus our attention on the two simplest cases
shown in Table~\ref{t:solutions}. For $F(g)=V_0$, for which the scalar
potential is of the exponential form, the \textit{exact} solution of
the WDW equation reads
\begin{subequations}
\begin{eqnarray}
\Psi &=& \exp \left[ \frac{k}{2} \left(
  \frac{\alpha}{3}-\frac{\phi}{B} \right) + \frac{Q}{2} \alpha + \frac{\mu^2
  V_0}{2B}\phi -\frac{1}{\mu}e^{3\alpha-B \phi} \right] \, , \label{expowf} \\
k &=& -3\left[ 3 \pm B \sqrt{1+Q^2/(\mu^2 V_0)} \right] \, ,
\end{eqnarray}
\end{subequations}
where the last equation is the solution to the constraint~(\ref{wdwho}).

\begin{figure}
\includegraphics[width=8cm]{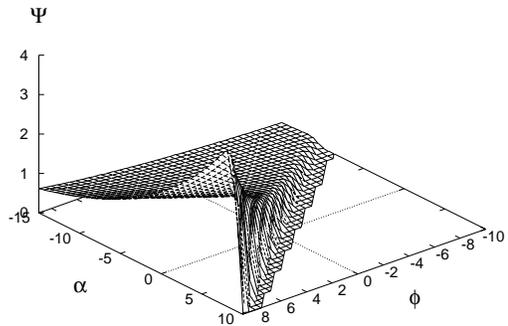}
\caption{\label{fig:esc1} Exact wave function~(\ref{expowf}) for the
  exponential potential shown in Table~\ref{t:solutions} with $\mu^2
  V_0 =1$ and $Q=0$. For this case, $k>0$ and then the wave function
  is different from zero only in a certain region of the
  ${\alpha,\phi}$ plane. We notice a sharp decay of the wave function
  in the region $3\alpha -B\phi >0$ due to exponential behavior of the
  superpotential function $S$ in Eq.~(\ref{expowf}).}
\end{figure}

One needs a not increasing wave function in order to recover classical
solutions at late times. Therefore, the wave function should be well
behaved for $|\alpha| \rightarrow \infty$ for fixed $\phi$. These two
conditions are accomplished if $k>0$; which in turn implies that
$\mu^2 V_0 >0$. Such a wave function is shown in Fig.~\ref{fig:esc1},
where we see that it is different from zero only in a certain region of
the ${\alpha,\phi}$ plane. Moreover, the wave function has a very well
definite boundary line, and this would tell us that the region for
which $\phi < (3/B) \alpha$ would not be allowed for the evolution of
the universe.

It is interesting to note that the scalar potential of the form $V\sim
e^{-2B \phi}$, where $B$ is an arbitrary parameter, is one
of the most studied in the literature, for basic literature see
\cite{Ratra:1987rm,Chimento:1995da,Copeland:1997et,Lyth:2000,Peebles:2002gy,Alam:2004jy,Sen:2002an,expo1}
and references therein. Such an exponential potential, in a scalar
field dominated universe, is inflationary if $2B <
\sqrt{2}$~\cite{Chimento:1995da,Copeland:1997et,expo1}. This regime is
reached if $-18 < 2 \mu^2 V_0 < -17$; hence $\mu$ is a purely imaginary
number and $B < 3$. Unfortunately, we cannot have a well behaved wave
function even in the case of $Q\neq 0$.

The other interesting case we are to take into account is $F(g)=1/g$,
for which functions $S$ and $W$ read
\begin{subequations}
\label{sinh}
\begin{eqnarray}
S &=& \frac{\mu}{9} e^{3\alpha} \sinh^2 (3\phi /2) \, , \label{sinha} \\
W &=& e^{Q\alpha/2} \left[ \cosh \left( 3\phi/2 \right) e^\alpha
  \right]^{k/6} \left| \tanh\left( 3\phi/2 \right) \right|^{-1} \,
. \label{sinhb}
\end{eqnarray}
\end{subequations}
Strictly speaking, taking $k=\textrm{const.}$ does not satisfy
Eq.~(\ref{wdwho2}). However, it can be shown that the latter can be
approximately satisfied if we are far from the line $\phi=0$ and $|k|
\ll 1$.

\begin{figure}
\includegraphics[width=8cm]{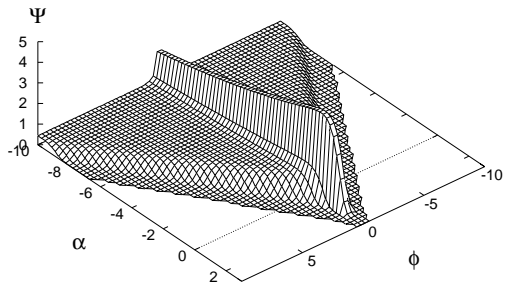}
\caption{\label{fig:esc2} Semiclassical wave function for a sinh
  scalar potential, which is the second solution shown in
  Table~\ref{t:solutions}. Here, $k=0.5$, $\mu=1$ and $Q=0$. Note that the
  wave function is limited by two well-definite boundary lines; it
  vanishes in the region in which $S>1$. However, there is a
  divergence in the line $\phi=0$, in which solution~(\ref{sinh}) does
  not comply with the constraint equation~(\ref{wdwho}), see text for
  details.}
\end{figure}

The wave function corresponding to Eqs.~(\ref{sinh}) is shown in
Fig.~\ref{fig:esc2}. Notice that the wave function diverges near the
line $\phi=0$, where solution~(\ref{sinh}) is not valid; but the wave
function is still constrained by two well definite boundary
lines. These lines arise from the superpotential
function~(\ref{sinha}), and represent the confinement of the wave
function due to the scalar field potential.

It is interesting to note that the wave function only \textit{goes} up
to $\alpha=0$. The reason for this can be understood in terms of the
classical solution: the universe reaches the origin of coordinates in
an infinite cosmic time $t$, see Eq.~(\ref{classicalb}). Although the
scalar potential has a minimum at $\phi=0$, the classical solution
does not show any oscillatory behavior around such a minimum.

We have seen how using canonical quantization for cosmological models
gives an equation whose solutions fix the form of the potential, and
then a family of potentials that are valid at the semiclassical level
was found. We believe that this method could be applied to other
different models (tachyon cosmology for instance) to find the form of
the potential, or as a toy model for the string theory landscape.   

\acknowledgments{This work was partially supported by CONACYT grant
  42748, CONCYTEG 05-16-K117-032, and PROMEP grant UGTO-CA-3}

\bibliography{escalarefs}

\end{document}